# Network on Chip: a New Approach of QoS Metric Modeling Based on Calculus Theory.

Salem NASRI [1,2]

[1] CES Lab , ENIS, Sfax, Tunisia.
salem.nasri@enim.rnu.tn
[2] College of Computer, Qassim university,
Kingdom of Saudi Arabia
snasri@qu.edu.sa

*Abstract:*

*According to ITRS, in 2018, ICs will be able to integrate billions of transistors, with feature sizes around 18 nm and clock frequencies near to 10 GHz. In this context, Network on Chip (NoC) appears as an attractive solution to implement future high performance networks and more QoS management. A NoC is composed by IP cores (Intellectual Propriety) and switches connected among themselves by communication channels. End-to-End Delay (EED) communication is accomplished by the exchange of data among IP cores. Often, the structure of particular messages is not adequate for the communication purposes. This leads to the concept of packet switching. In the context of NoCs, packets are composed by header, payload, and trailer. Packets are divided into small pieces called Flits. It appears of importance, to meet the required performance in NoC hardware resources. It should be specified in an earlier step of the system design. The main attention should be given to the choice of some network parameters such as the physical buffer size in the node. The EED and packet loss are some of the critical QoS metrics. Some real-time and multimedia applications bound up these parameters and require specific hardware resources and particular management approaches in the NoC switch.*

*A traffic contract (SLA, Service Level Agreement) specifies the ability of a network or protocol to give guaranteed performance, throughput or latency bounds based on mutually agreed measures, usually by prioritizing traffic. A defined Quality of Service (QoS) may be required for some types of network real time traffic or multimedia applications.*

*The main goal of this paper is, using the Network on Chip modeling architecture, to define a QoS metric. We focus on the network delay bound and packet losses. This approach is based on the Network Calculus theory, a mathematical model to represent the data flows behavior between IPs interconnected over NoC.*

*We propose an approach of QoS-metric based on QoS-parameter prioritization factors for multi applications-service using calculus model.*

## 1. INTRODUCTION

With the rapid development of advanced technology and high speed communication systems, the quality of service (QoS) is becoming one of the most important aspects of networks. The concept of QoS concerns the classes of services for applications offered by a network. Hence in order to evaluate the efficiency of the network a QoS metric is needed.

In this work, we address the QoS Metric problem for NoC based system. We focus on the study of the delay introduced by the switcher in a mesh topology.

We assume for this study the following characteristics: The modules are interconnected by a network of multi-port switches connected to each other by links made of parallel point-to-point lines. The network applies a mesh topology using the X-Y routing algorithm with input queuing.

This paper is organized as follows: Section 2 presents network architecture overview; Section 3 describes the network modeling based on *Network Calculus* theory, in section 4 we present a QoS modeling based on QoS prioritization parameters. The conclusion outlines the main contribution of this work and its future issues.

## 2. NoC Architecture Overview

The target NoC on chip architecture is a scalable packet switched communication platform for single chip systems. The NoC architecture consists of a (nxn) mesh of switches interconnecting resources. Figure 1 shows a NoC architecture with 16 resources. Each switch is connected to its neighbors and one resource. Switches have two, three or four bidirectional links with neighbors depending on the position of each in the graph. Resources are heterogeneous. It can be a processor core, a memory block, an FPGA, a custom hardware block or any other intellectual property (IP) block, which fits into the available slot and complies with the interface with the NoC switch. We assume that switches in NoC have buffers to manage data traffic. Every resource had one specific address and is connected to a switch in the network over a Resource Network Interface (RNI). The NoC architecture defines four protocol layers: the physical layer, the data link layer, the network layer, and the transport layer [2, 5, 6, 10].

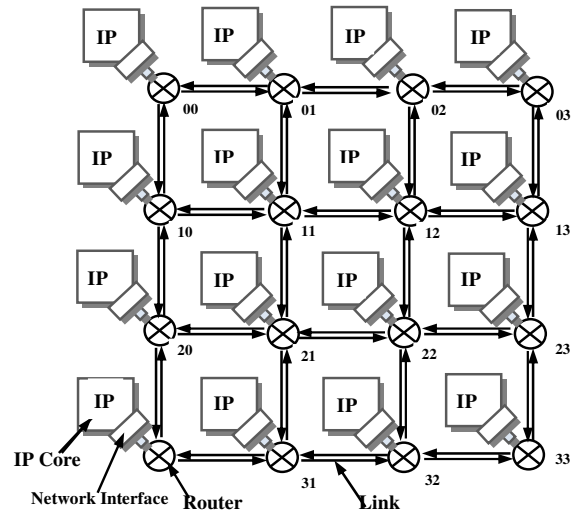

Fig 1: Target NoC architecture

## 3. NoC Architecture Modelling
### 3.1.    Network calculus

Network calculus is a general paradigm used for data flow modeling in communication process. It provides provision for QoS in communication networks [3, 7, 9]. The main principle of network calculus is to show that if all the input flows to a network satisfy a certain set of constraints, then so do all the flows within the network [9]. The formulation of the constraints is enough easy to allow the computation of bounds on various performance measures, such as delay and queue length at each element of the network [4].

A well-known Network Calculus is the (σ, ρ) calculus, first introduced by R. Cruz in [7] and further developed in [8]. It provides deterministic bounds on delay and buffering requirements in a communication network. This approach is useful for modelling applications requiring deterministic QoS guarantees.

Network calculus lies in the research of the most unfavourable situations, which makes it possible to obtain rising. For these considerations, various representations and concepts are introduced [1, 7, 8, 9].

Suppose we are given a stream of traffic flowing on a communication system used as input traffic link between two network nodes. This system can be represented by Leaky bucket traffic Controller Model, figure 2. A Leaky Bucket Controller is a device that analyzes the data on a flow R(t) as follows. There is a pool (bucket) of fluid of size σ. The bucket is initially empty. The bucket has a hole and leaks at a rate of ρ units of fluid per second when it is not empty.

We represent this traffic by a nonnegative function R(t) as follows: for any instant t, and for any

$y \geq x \in R$, $\int_x^y R(t)dt$ is the amount of data transmitted on the link in the interval [x, y].

Thus, in general, R(t) represents the instantaneous rate of traffic from the stream flowing on the node at time t. We often say that R is the rate function of the stream. The study of the switcher time processing of the backlog makes possible to bound up the latency time in NoC switch.

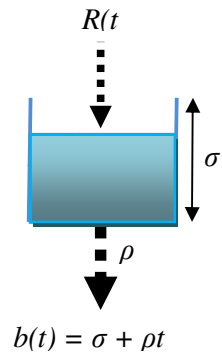

$$b(t) = \sigma + \rho t$$

Fig 2: Leaky bucket Controller Model

Data from the flow R(t) has to pour into the bucket an amount of fluid equal to the amount of data. Data that would cause the bucket to overflow is declared non-conformant; otherwise the data is declared conformant.

The bucket fluid size σ and the quantity of data in gust and the rate of fluid ρ are according by the following equation:

$$b(t) = \sigma + \rho t \qquad (1)$$

Then we can write:

$$R(t) \approx b(t) \Leftrightarrow \int_x^y R(t)dt < \sigma + \rho(y - x) \qquad (2)$$

## 3.2. The FIFO Queue:

The FIFO queue can be viewed as a degenerate form of a FIFO multiplexer. The FIFO has one input link and one output link. The input link has transmission capacity $C_{in}$, and the output link has transmission capacity $C_{out}$. The FIFO is defined simply as follows. Data that arrives on the input link is transmitted on the output link in FCFS (First Come First Served) order as soon as possible at the transmission rate $C_{out}$. For example, if a packet begins to arrive at time $t_0$ and if no backlog exists inside the FIFO at time $t_0$, then the packet also commences transmission on the output link at time $t_0$. We assume that $C_{in} \geq C_{out}$, so that this is possible; if $C_{in}$ were less than $C_{out}$, then this would be impossible to do-the FIFO would "run out" of data to transmit immediately following time $t_0$, before the packet could be entirely transmitted at rate $C_{out}$. Note that this assumption is always satisfied if the data passes through a receive buffer immediately prior to entering the FIFO. The size of the backlog inside the FIFO at time t and is given by:

$$W_{C_{out}}(R)(t) = \max_{s \leq t}[\int_s^t R - C(t-s)] \tag{3}$$

Note that the j[th] packet, which arrives at times $S_j$, must wait for all of the current backlog and this backlog gets transmitted at rate $C_{out}$. It follows that the j[th] packet commences exit from the FIFO at time tj = $S_j$ + $d_j$, where:

$$d_j = \frac{1}{C_{out}} W_{C_{out}}(R_0)(t)$$

$$R_0 \approx b(t) \Leftrightarrow \int_s^t R_0(t)dt < b(t-s) \text{ and } \int_s^t R_0(t)dt < \sigma + \rho(t-s) \tag{4}$$

$$d_j(s_j) = \frac{1}{C_{out}}W_{C_{out}}(R_0)(t) = \frac{1}{C_{out}}[\int_s^{s_j} R_0 - C_{out}(s_j-s] \leq$$

$$\frac{1}{C_{out}}[\sigma_{in} + \rho_{in}(s_j-s) - C_{out}(s_j-s)] \leq \tag{5}$$

$$\frac{1}{C_{out}}\max_{s \leq s_j}[\sigma_{in} - (C_{out}-\rho_{in})(s_j-s)] = d_j(s_j)$$

*S: is a maximum argument for $W_{Cout}(R_0)(t)$.*

Then, *the delay of any data bit entering an FIFO from input link is upper bounded by*:

$$D_{queue} = \frac{1}{C_{out}}[\sigma_{in} - \frac{C_{out}-\rho_{in}}{C_{in}}L] \tag{6}$$

### 3.3. The Demultiplexer (Demux)

The demultiplexer (Demux) has a single input link and two or more output links. The function of the DEMUX is to "split up" two or more substreams that arrive on the input link and route the substreams to the appropriate output link. It is assumed that the data is "marked" so that the Demux can instantaneously determine which substream any given packet arriving on the input link belongs to; recall that we assume a fixed routing discipline.

### 3.4. The multiplexer (Mux)

The multiplexer has two or more input links and a single output link. As the name suggests, the function of the Mux is to merge the streams arriving on the input links onto the output link.

For simplicity of exposition, we explicitly analyze multiplexers with only 2 input links. We assume that the first input link has maximum transmission rate $C_1$ and that the second input link has maximum transmission rate $C_2$. The output link has maximum transmission rate $C_{out}$ and we assume $C_i \geq C_{out}$ for i = 1,2. It is necessary to make this assumption so that it is possible for a given data packet to start transmission at rate $C_{out}$ on the output link at the same time it begins to arrive to the multiplexer; this is known as "cut-through" switching. In a practical sense, there is no loss, in general, in making this assumption: if the data passes through receive buffers immediately prior to entering the multiplexer, then the input links to the multiplexer will have an infinite transmission rate and the assumption is satisfied.

Let the rate of the input stream arriving on the first input link be represented by R1; similarly, let the rate of the second input stream be represented by R2. Let $R_{1,out}$ and $R_{2,out}$ represent the rate of the first and second streams, respectively, as they exit the multiplexer. Thus, for example, $R_{out} = R_{1,out} + R_{2,out}$ represents the rate of the traffic stream that appears on the output link of the multiplexer. Finally, define $b_{i(t)}$ as the size of the backlog from input stream i that exists in the multiplexer at time t, and let b(t) = $b_1(t)$+ $b_2(t)$.

We consider the first-come first-served multiplexer (FIFO Mux). In the FIFO Mux, packets are transmitted on the output link in the order in which they arrive on the input links. "Ties" are broken arbitrarily. By definition, a packet arrives at the instant it first begins transmission on the input link. We assume that the FIFO Mux is work-conserving, i.e., if b (t) > 0, then $R_{out}$ (t) = $C_{out}$

In these considerations the delay of any data bit entering an FIFO multiplexer from stream is upper bounded by the expression of $D_{mux}$ [7, 8, 12].

$$\bullet \quad D_{mux} = \frac{1}{C_{out}} \max_{t \geq 0} [b_1(t) + b_2(t + \frac{L}{C_2}) - C_{out}(t)] \tag{7}$$

The total delay of any data entering on the NoC switcher architecture is then upper bounded by:

$$D_{NoC} = D_{Mux} + D_{queue} \tag{8}$$

The delay given is invariant of time, takes again the technological parameters of the multiplexing and the queuing and depends only on two anonymous factors, $\sigma$ and $\rho$, the parameters of the curves of arrivals of the various traffics.

## 4. QoS Modeling and Measurements

A traffic contract (SLA, Service Level Agreement) specifies the ability of a network or protocol to give guaranteed performance, throughput or latency bounds based on mutually agreed measures, usually by prioritizing traffic. A defined Quality of Service may be required for some types of network real time traffic or multimedia application [11, 13, 14]. We propose an approach of QoS-metric based on QoS-parameter prioritization factors $\alpha_i$ for one application-service using the relation:

$$Q(p_1, p_2, p_3, \ldots, p_m) = F(\alpha_i p_i), \qquad i = 1, \ldots, m \tag{9}$$

We define $k$, $\alpha_i$, $p_i$, and $Q(p_1, p_2, p_3, \ldots, p_m)$ such as:

1-    $k \geq 1$: network efficiency coefficient ( in our case we chose k= 1.1 for example).

2-    $\alpha_i$ : parameter prioritization factor, with:

$$\sum_{i=1}^{m} (\alpha_i) = 1 \tag{10}$$

3-    $p_i$ : QoS performance parameter,        $p_i$ should be normalized $p_{in}$
$p_{imax} = Max\{ p_i \}$, $p_{imin} = Min\{ p_i \}$,

   a.    For increasing parameters when buffer size increases

$$p_{in} = \left| \frac{p_i - p_{i\min}}{k*(p_{i\max} - p_{i\min})} \right| \tag{11}$$

   b.    For decreasing parameters when buffer size increases

$$p_{in} = \left| \frac{p_{i\max} - p_i}{k*(p_{i\max} - p_{i\min})} \right| \tag{12}$$

4-    Then the QoS expression can be defined by:

$$Q(p_1, p_2, \ldots, p_m) = \sum_{i=1}^{m} (\alpha_i p_{in}) \tag{13}$$

In this model we consider the packet loss parameter as $p_1$ and the EED parameter as $p_2$ and FIFO scheduling techniques for using different buffer sizes, $\alpha_1$, $\alpha_2$ are arbitrarily fixed referring to the equation (10).

## 5. Experimentation and Results

The studied architecture is a 4x4 Mesh. The maximum bandwidth link is fixed to 2GB/s. The purpose of the study is to optimise buffer size according to the general network loading states and also according to the interconnected IPs throughput.

The traffic is transferred over the network between tow IPs interconnected to the 00 switch (source) and the thirty third switch. The packet size is of 4 bytes (32 bits) based on a by 8 bits flits (4 flits by packet).

## 5.1. Application rates and data losses

A reasonable buffer size may drive on network on chip performances. The main idea is to keep the minimum buffer size with avoiding packet losses and over end to end delays. Otherwise, the hardware cost will be large. It may influence both network quality of service performances and hardware designing resources optimization.

We mainly focus on per cent flit losses due to buffer congestion for a network loading. This helps us to identify the optimal buffer size for the switch design. Figure 5 shows the loosed data increases with low size of buffer and application rate.

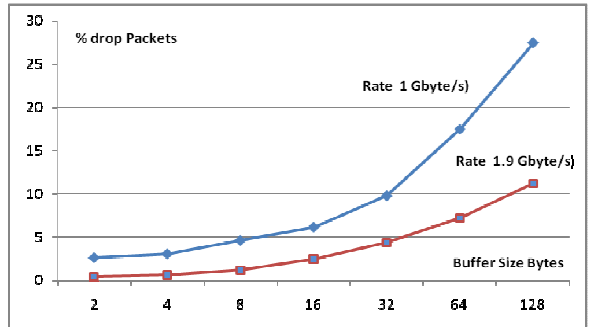

Figure 4. % dropped packets and application rate

Figure 4 shows the relationship between % packet dropped and the available switch buffer size. It carries out boundary values of buffer size for different rates and also their associated per cent packet dropped. This figure shows that the % dropped packets decreases when application rate increases. We can say that when the buffer size decreases the performance of the switch increases. That is normal in our sense because a high buffer size contribute on the switch congestion. A buffer size of 16 or 32 bytes leads to a better performance of the switch.

## 5.2. End to end delay and buffer size

The end to end delay is one of the major critical QoS metrics. Some real-time applications bound up this value and require a specific hardware resources and particular management approaches in the NoC switch.

Figure 5 sums up the end to end delay when the switching buffer is managed by FIFO scheduling approach. This figure shows that the end to end delay decreases when the application rate increases with the buffer size increasing.

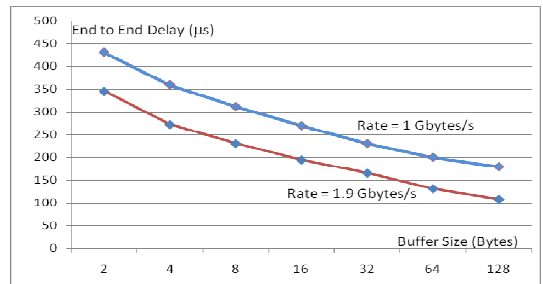

Figure 5. End to end delay and buffer size

## 5.3. QoS measurements and analysis

Combining different values of parameter prioritization factor the QoS becomes sensible for these combinations. Figure 6 and 7 show the sensibility of the QoS to different parameters combinations. We can see also that the buffer size is a critical QoS parameter.

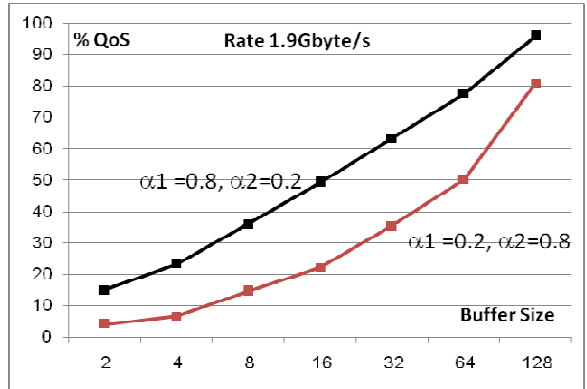

Figure 6. %QoS and buffer size for $\alpha_1 \# \alpha_2$ and Application Rate =1.9 Gbyte/s

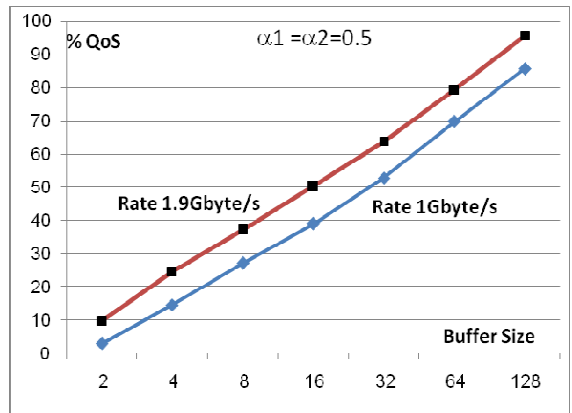

Figure 7. %QoS and buffer size $\alpha_1 = \alpha_2$ and different application Rates

# 6. CONCLUSION AND FUTURE WORKS

This paper studies the NoC switcher modelling with network calculus theory. It presents a model providing the bounds for network delay and buffering requirements.

The main goal of this work is a contribution to network flow modeling with theoretical approach. As a result, a model for network delay is developed. It contributes on the QoS evaluation and NoC management in where critical time applications are enhanced.

This work has completed with a study and development of a QoS model of multi application systems with multi parameters. This helps to make up the efficiency of the QoS metric evaluation.

**Pr. Salem Nasri**

Received his PhD in Automatic Control and Computer Engineering from 'INSA: Institut National des Sciences Appliquées' Toulouse, France, in June 1985. He obtained the diploma of "HDR: Habilitation à Diriger les Recherches" in Computer Engineering, in May 2001 from the 'ENIS: Ecole Nationale d'Ingénieurs de Sfax', Tunisia. He is simultaneously Professor at 'ENIM: Ecole Nationale d'Ingénieurs de Monastir', Tunisia, and at the Computer College, Qassim University, Kingdom of Saudi Arabia. His research interests are: Computer Networks, Network on chip, High Speed Protocols, Wireless Communication Systems, Multimedia Applications and Quality of Service Modeling.

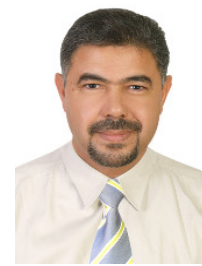